%Paper: hep-th/9502044
%From: VITIELLO@VAXSA.DIA.UNISA.IT (G.Vitiello, tel. +39 89 822311 - fax +39 89
%%953804)
%Date: Tue, 7 Feb 1995 12:31:51 +0100 (WET)

% This file is written in TEX.

% Y.N.Srivastava, G.Vitiello and A. Widom
% QUANTUM DISSIPATION AND QUANTUM NOISE
% To be published in Annals of Phys. (N.Y.) 238, 200 (1995)

\magnification 1200
\overfullrule 0 pt
\font\abs=cmr9
\font\ist=cmr8

\font\uit=cmu10

\def\CcC{{\hbox{\tenrm C\kern-.45em{\vrule height.65em width0.07em depth-.04em
\hskip.45em }}}}
\def\RrR{{\hbox{\tenrm I\kern-.17em{R}}}}
\def\HhH{{\hbox{\tenrm {I\kern-.18em{H}}\kern-.18em{I}}}}
\def\DdD{{\hbox{\tenrm {I\kern-.18em{D}}\kern-.36em {\vrule height.62em
width0.08em depth-.04em\hskip.36em}}}}
\def\ZzZ{{\hbox{\tenrm Z\kern-.31em{Z}}}}
\def\IiI{{\hbox{\tenrm I\kern-.19em{I}}}}
\def\NnN{{\hbox{\tenrm {I\kern-.18em{N}}\kern-.18em{I}}}}
\def\QqQ{{\hbox{\tenrm {{Q\kern-.54em{\vrule height.61em width0.05em
depth-.04em}\hskip.54em}\kern-.34em{\vrule height.59em width0.05em
depth-.04em}}
\hskip.34em}}}
\def\OoO{{\hbox{\tenrm {{O\kern-.54em{\vrule height.61em width0.05em
depth-.04em}\hskip.54em}\kern-.34em{\vrule height.59em width0.05em
depth-.04em}}
\hskip.34em}}}

\def\uq2{U_q({\uit su}(2))}

\def\fraz#1#2{{\strut\displaystyle #1\over\displaystyle #2}}

\def\part#1{\fraz{\partial}{\partial#1}}

\def\su2q{SU(2)_q}
\def\h1q{H(1)_q}

\def\nu{N_{1}}

\hsize= 15 truecm
\vsize= 22 truecm
\hoffset= 0.5 truecm
\voffset= 0 truecm

\null\vskip1.5truecm

\baselineskip= 13.75 pt
\footline={\hss\tenrm\folio\hss}
%\nopagenumbers
\centerline
{\bf  QUANTUM DISSIPATION AND QUANTUM NOISE}
\bigskip
\centerline{
{\it    Y.N.Srivastava${}^{+,\dagger}$, G.Vitiello${}^*$ and
A.Widom${}^{+,\dagger}$}}
\bigskip
{\ist ${}^+$Physics Department, Northeastern University, Boston, MA 02115,
USA}

{\ist ${}^\dagger$Dipartimento di Fisica, I.N.F.N.,
Universit\`a di Perugia, I06100 Perugia,
Italy}
\footnote{}{\hskip -.85truecm {\abs E--mail: srivastava@pg.infn.it
{}~ vitiello@sa.infn.it  ~44192::widom
 \hfill}}

{\ist ${}^*$Dipartimento di Fisica, I.N.F.N., Universit\`a di Salerno,
I84100 Salerno, Italy}

\bigskip
\bigskip
\bigskip
\bigskip
\bigskip
\bigskip
\bigskip

\noindent{\bf Abstract.} {\abs We derive the exact action for a damped
mechanical system ( and the special case of the linear oscillator)
from the path integral formulation of the quantum Brownian motion problem
developed by Schwinger and by Feynman and Vernon. The doubling of
the phase-space degrees of freedom for dissipative systems and
thermal field theories is discussed and the initial values of
the doubled variables are related to quantum noise effects.}

\noindent
PACS  05.70.Ln; 11.10.Ef; 42.50.+q;

\vfill
\eject

\bigskip
\centerline {\bf 1. Introduction}
\bigskip

A dissipative system is physically incomplete and a microscopic theory
must include the details of processes responsible for dissipation,
including quantum
effects. One then would start from the beginning with a Hamiltonian that
describes a complete system, the bath and the system-bath interaction.
Subsequently, the description of the original dissipative system is
recovered by the reduced density matrix obtained by eliminating
the bath variables which originate the damping and the fluctuations.
The problem with dissipative systems in quantum mechanics is indeed that
canonical commutation relations (ccr) are not preserved by time evolution
due to damping terms. The role of fluctuating forces is in fact the one
of preserving the canonical structure.

The above strategy, however,
may not always be viable since it requires the knowledge of
the details of the processes inducing the dissipation;
these details may not be explicitely known and the dissipation
mechanisms are sometime globally described by such parameters as friction,
resistance, viscosity etc.. In such a case a different route must be
followed. On the other hand, the attempt to derive, from a variational
principle, the equations of motion defining the dissipative system
requires the introduction of additional complementary
equations[1].

The latter approach has been pursued in refs. 2 and 3 where
the quantization of the one-dimesional
damped linear harmonic oscillator (dho) has been studied by doubling
the phase-space degrees of freedom. The new degrees of freedom
thus introduced play the role of the bath degrees of freedom and make
it possible to derive the dho equation from a variational principle.

The quantum hamiltonian is obtained as
$$
\eqalign{
{\cal H} &= {\cal H}_{o} + {\cal H}_{I} \quad , \cr
{\cal H}_{o} = \hbar \Omega ( A^{\dagger} A - B^{\dagger} B ) \quad &,
\quad
{\cal H}_{I} = i \hbar \Gamma ( A^{\dagger} B^{\dagger} - A B ) \quad ,
\cr}
\eqno(1)
$$
where $\Gamma $
is the decay constant, $\Omega$ the
frequency and $ A^{\dagger}$, $B^{\dagger}$, $A$ and $B$ are creation
and annihilation operators with usual ccr[2,3].

As pointed out in refs.
2 and 3, only if one restricts the Hilbert space
to the subspace of
those states, $|\psi >$, which are annihilated by $B$: $B |\psi >
= 0$, the eigenstates of ${\cal H}$ merge into those of the simple
undamped harmonic oscillator when $\Gamma \rightarrow 0$.
Therefore the
states generated by $B^{\dagger}$ represent the sink where the energy
dissipated by the quantum damped oscillator flows:
the $B$-oscillator thus
represents the reservoir or heat bath coupled to the
$A$-oscillator.

The
dynamical group structure is that of $SU(1,1)$ and
it has also been shown[3] that the canonical quantization of
the dho can be achieved in a consistent way in the infinite volume
(or thermodynamic) limit of Quantum Mechanics, i.e. in
Quantum Field Theory (QFT) where infinitely many unitarily inequivalent
representations of the ccr are allowed. The reason for this is that
the set of states of the dho splits into unitarily inequivalent
representations (i.e. into disjoint {\it folia}, in the ${\cal C}^{*}$
-algebra formalism) each one representing the states of the oscillator
at time $t$ and therefore QFT provides the proper setting
to discuss dho: the irreversible, non-unitary time evolution is thus
described as {\it tunneling}
across unitarily inequivalent representations
(breakdown of time reversal invariance or
{\it arrow of time}). In such a description a central role is
played by the degeneracy among the {\it vacua} of the inequivalent
representations, which is formally expressed by the fact that
the free Hamiltonian ${\cal H}_{o}$
is the Casimir
operator of the $SU(1,1)$ group and therefore a constant of motion,
$[\, {\cal H}_{o} , {\cal H}_{I}\, ] = 0$. Quantum processes
changing separately $n_{A}$ and
$n_{B}$ are allowed provided their
difference is left constant and therefore ${\cal H}_{o}$ may have
(infinitely) many degenerate zero-energy states (vacua).
The positiveness of ${\cal H}_{o}$ is also ensured by its constancy
in time, once its initial (positive) value has been assigned.

In ref. 3 it has been shown that the dho states are time dependent
thermal states, as expected due to the statistical nature of dissipation.
In particular the formalism
for the dho turns out to be similar to the one of real time
QFT at finite temperature,
also called thermo-field dynamics (TFD)[4], which is indeed based on
the doubling of the degrees of freedom.

The purpose of this paper is the following.
In the approach
of refs.1-3 the system is perturbed in a manner that its time
development is governed by different
dynamics depending on the positive and negative sense of time. In this
work we want to show that there is
consistency between the variational
principle in the Lagrangian formalism and such different dynamical
behaviors relative to a different sense of time. Or, in other words,
on a closed time path the variational principle is sufficient to
determine the expectation values of any physical observable for a
given initial condition.
We will obtain the exact action for the dho in the path integral
formalism. In particular we will reach some conclusion
on the initial values of the doubled variables and relate them
to the probability of quantum fluctuations in the ground state,
a result which is interesting also in the more general case
of thermal field theories, such as TFD, with the doubled variable
formalism.

We will cast the canonical quantization scheme of the dho of
ref. 3 in the formalism developed
by Schwinger[5] and Feynman and Vernon[6] since it is particularly
suited to our task.

\bigskip
\bigskip
\centerline{\bf 2. Lagrangian formalism }
\bigskip

Our purpose is to explore the manner in which the Lagrangian model for
quantum dissipation of refs. 2-3 arises
from the formulation of the quantum Brownian
motion problem as described by Schwinger[5] and by Feynman and Vernon[6].
Let us first recall (and generalize) the Lagrangian of dho[2,3].
The Langrangian for a particle of mass $\mu $, damped by a mechanical
resistance $R$ moving in a potential $V$, reads
$$
{\cal L}(\dot{x},\dot{y},x,y)=\mu \dot{x}\dot{y}
-V(x+(1/2)y)+V(x-(1/2)y)+(R/2)(x\dot{y}-y\dot{x}). \eqno(2)
$$
The classical Lagrangian equations of motion implied by Eq.(2) read
$$
\mu \ddot{x}+R\dot{x}+
(1/2)[V^\prime (x+(1/2)y)+V^\prime (x-(1/2)y)]=0,  \eqno(3a)
$$
$$
\mu \ddot{y}-R\dot{y}+
[V^\prime (x+(1/2)y)-V^\prime (x-(1/2)y)]=0.  \eqno(3b)
$$

It is easy to see that for $V(x \pm (1/2)y) = (1/2)k(x \pm (1/2)y)^{2}$ Eqs.
(3) give the dho equation and its complementary equation of refs. 2 and 3.

If from the manifold of solutions to Eqs.(3) we choose those for which
the $y$ coordinate is constrained to be zero, then Eqs.(3) simplify to
$$
y=0;\ \ \mu \ddot{x}+R\dot{x}+V^\prime (x)=0. \eqno(3c)
$$
Thus we obtain a classical damped equation of motion from a Lagrangian
theory at the expense of introducing an ``extra'' coordinate $y$, later
constrained to vanish. (Note that $y(t)=0$ is a true solution to
Eqs.(2) so that the constraint is {\it not} in violation of the equations
of motion.)

The question to which we address ourselves is the following.
Does the introduction of an ``extra coordinate'' make any sense in the
context of conventional quantum mechanics? Consider the special case
of zero mechanical resistance. For that case, one should begin with the
Hamiltonian for an isolated particle,
$$
H=-(\hbar^2/2\mu )(\partial/\partial Q)^2 +V(Q). \eqno(4)
$$
For an isolated particle one obtains in quantum mechanical theory the
density matrix equation of motion
$$
i\hbar (\partial \rho /\partial t)=[H,\rho ], \eqno(5)
$$
which indeed requires two coordinates (say $Q_+$ and $Q_-$). In the
coordinate representation, Eqs.(4) and (5) read
$$
i\hbar (\partial/\partial t)<Q_+|\rho (t)|Q_->=
$$
$$
\{
-(\hbar^2/2\mu)[(\partial/\partial Q_+)^2-(\partial/\partial Q_-)^2]
+[V(Q_+)-V(Q_-)]
\}<Q_+|\rho (t)|Q_->. \eqno(6)
$$
Employing the coordinates $x$ and $y$,
$$
Q_{\pm}=x\pm (1/2)y, \eqno(7)
$$
and the associated density matrix function
$$
W(x,y,t)=<x+(1/2)y|\rho (t)|x-(1/2)y>, \eqno(8)
$$
simplifies Eq.(6) which now reads
$$
i\hbar (\partial /\partial t) W={\cal H}_oW \eqno(9a)
$$
$$
{\cal H}_o=(p_xp_y/\mu)+V(x+(1/2)y)-V(x-(1/2)y), \eqno(9b)
$$
$$
p_x=-i\hbar(\partial /\partial x),
\ \ p_y=-i\hbar(\partial /\partial y). \eqno(9c)
$$
Of course the ``Hamiltonian'' Eq.(9b) may be constructed from the
``Lagrangian''
$$
{\cal L}_o=\mu \dot{x}\dot{y}-V(x+(1/2)y)+V(x-(1/2)y). \eqno(10)
$$
We have then the justification for introducing Eq.(2)
at least for the case $R=0$.

Finally we observe that for $V(x \pm (1/2)y) =
(1/2)k(x \pm (1/2)y)^{2}$ it is easy to obtain the Hamiltonian (1)
from the Lagrangian Eq.(2)[2,3].

We also notice that ${\cal H}_{o}$ and $\cal H_{I}$ in Eq.(1)
are the free Hamiltonian and the generator of Bogolubov
transformations,respectively, in TFD[4]. Our present discussion
thus includes the doubling of degrees of freedom in finite temperature
QFT.

\bigskip
\bigskip
\centerline {\bf 3. Damping}
\bigskip

Now let us suppose that the particle interacts with a thermal bath
at temperature $T$. The interaction Hamiltonian between the bath and
the particle is taken as
$$
H_{int}=-fQ, \eqno(11)
$$
where $Q$ is the particle coordinate and $f$ is the random force on
the particle due to the bath.

In the Feynman-Vernon[6] quantum Brownian motion scheme, the effective
action for the particle has the form
$$
{\cal A}[x,y]=\int_{t_i}^{t_f}dt{\cal L}_o(\dot{x},\dot{y},x,y)
+{\cal I}[x,y], \eqno(12)
$$
where ${\cal L}_o$ is defined in Eq.(10) and
$$
e^{(i/\hbar){\cal I}[x,y]}=
<(e^{(-i/\hbar)\int_{t_i}^{t_f}f(t)Q_-(t)dt)})_-
(e^{(i/\hbar)\int_{t_i}^{t_f}f(t)Q_+(t)dt)})_+>. \eqno(13)
$$
In Eq.(13): (i)The average is with respect to the thermal bath; (ii)
``$(.)_{+}$'' denotes time ordering and ``$(.)_{-}$''
denotes anti-time ordering; (iii) The c-number coordinates $Q_{\pm }$
are defined as in Eq.(7).
We observe that
if the interaction between the bath and the coordinate $Q$ (i.e
$H_{int}=-fQ$ ) were turned off, then the operator $f$ of the bath
would develop in time according to
$f(t)=e^{iH_R t/\hbar}fe^{-iH_R t/\hbar }$ where $H_R$ is the Hamiltonian
of the isolated bath (decoupled from the coordinate $Q$).
$f(t)$ is the force operator of the bath to be used in Eq.(13).

Assuming that the particle first makes contact with the bath at the
initial time $t_i$, the reduced density matrix function in Eq.(8)
obeys at a final time
$$
W(x_f,y_f,t_f)=\int_{-\infty}^{\infty }dx_i\int_{-\infty}^{\infty }dy_i
K(x_f,y_f,t_f;x_i,y_i,t_i)W(x_i,y_i,t_i), \eqno(14)
$$
with the path integral representation for the time development
$$
K(x_f,y_f,t_f;x_i,y_i,t_i)=
\int_{x(t_i)=x_i}^{x(t_f)=x_f}{\cal D}x(t)
\int_{y(t_i)=y_i}^{y(t_f)=y_f}{\cal D}y(t)
e^{(i/\hbar){\cal A}[x,y]}. \eqno(15)
$$

The evaluation of ${\cal I}[x,y]$ for a linear passive damping thermal
bath involves several Greens functions all of which have been discussed
by Schwinger[5] in his work on quantum Brownian motion. For completeness
of presentation we here discuss these results.

\bigskip
\bigskip
\centerline{\bf 4. Greens Functions}
\bigskip

The fundamental correlation function for the random force on the particle
due to the thermal bath is given by
$$
G(t-s)=(i/\hbar )<f(t)f(s)>.     \eqno(16)
$$
{}From this one may construct the time ordered Greens function and the
anti-time ordered Greens function
$$
G_+(t-s)=\theta (t-s)G(t-s)+\theta (s-t)G(s-t), \eqno(17a)
$$
$$
G_-(t-s)=\theta (s-t)G(t-s)+\theta(t-s)G(s-t),  \eqno(17b)
$$
where
$$
\theta(t-s)=1\ if\ t>s,  \eqno(18a)
$$
and
$$
\theta(t-s)=0\ if\ t<s. \eqno(18b)
$$

The retarded and advanced Greens functions are defined by
$$
G_{ret}(t-s)=\theta (t-s)[G(t-s)-G(s-t)], \eqno(19a)
$$
and
$$
G_{adv}(t-s)=\theta (s-t)[G(s-t)-G(t-s)]~. \eqno(19b)
$$

{}From a causal engineering view point, the mechanical impedance
$Z(\zeta )$ (analytic in the upper half complex frequency
plane ${\cal I}m
{}~\zeta >0$) is determined by the retarded Greens function
$$
-i\zeta Z(\zeta )=\int_0^\infty dt G_{ret}(t)e^{i\zeta t}. \eqno(20)
$$
The time domain quantum noise in the fluctuating random force
$$
N(t-s)=(1/2)<f(t)f(s)+f(s)f(t)>, \eqno(21)
$$
is distributed in the frequency domain
$$
N(t-s)=\int_0^\infty d\omega S_f (\omega)cos[\omega (t-s)], \eqno(22)
$$
in accordance with the Nyquist theorem
$$
S_f(\omega )=(\hbar \omega /\pi )
coth(\hbar \omega /2kT){\cal R}e Z(\omega +i0^+). \eqno(23)
$$
The mechanical resistance is defined
$$
R=lim_{\omega \rightarrow 0}{\cal R}e Z(\omega +i0^+). \eqno(24)
$$

Finally, the time ordered and anti-time ordered Greens functions
describe both the retarded and advanced Greens functions as well
as the quantum noise,
$$
G_\pm (t-s)=\pm (1/2)[G_{ret}(t-s)+G_{adv}(t-s)]
+(i/\hbar )N(t-s). \eqno(25)
$$

\bigskip
\bigskip
\centerline {\bf 5. Bath Interactions}
\bigskip

For the particle interacting with the bath describing a linear passive
mechanical impedance, Eq.(13) may be evaluated following Feynman and
Vernon as,
$$
{\cal I}[x,y]=(1/2)\int_{t_i}^{t_f}\int_{t_i}^{t_f}dtds
[G_+(t-s)Q_+(t)Q_+(s)$$
$$
+G_-(t-s)Q_-(t)Q_-(s)
-2G(t-s)Q_-(t)Q_+(s)]~. \eqno(26)
$$
Eq.(26) is more simple when expressed in terms of coordinates $x$ and $y$,
$$
{\cal I}[x,y]=(1/2)\int_{t_i}^{t_f}\int_{t_i}^{t_f}dtds
[G_{ret}(t-s)+G_{adv}(t-s)][x(t)y(s)+x(s)y(t)]
$$
$$
+(i/2\hbar )\int_{t_i}^{t_f}\int_{t_i}^{t_f}dtdsN(t-s)y(t)y(s). \eqno(27)
$$
Further simplification results if one defines the retarded force on $y$
$$
F_y^{ret}(t)=\int_{t_i}^{t_f}dsG_{ret}(t-s)y(s), \eqno(28a)
$$
and the advanced force on $x$,
$$
F_x^{adv}(t)=\int_{t_i}^{t_f}dsG_{adv}(t-s)x(s). \eqno(28b)
$$
The interaction between the bath and the particle is then
$$
{\cal I}[x,y]=(1/2)\int_{t_i}^{t_f}dt[x(t)F_y^{ret}(t)+y(t)F_x^{adv}(t)]
$$
$$
+(i/2\hbar )\int_{t_i}^{t_f}\int_{t_i}^{t_f}dtdsN(t-s)y(t)y(s). \eqno(29)
$$

\bigskip
\bigskip
\centerline {\bf 6. The Final Action}
\bigskip

Putting all the pieces together, we have from Eqs.(12) and (29)
the real part of the action
$$
{\cal R}e{\cal A}[x,y]=\int_{t_i}^{t_f}dt{\cal L}, \eqno(30a)
$$
$$
{\cal L}=\mu \dot{x}\dot{y}-[V(x+(1/2)y)-V(x-(1/2)y)]
+(1/2)[xF_y^{ret}+yF_x^{adv}], \eqno(30b)
$$
and the imaginary part of the action
$$
{\cal I}m{\cal A}[x,y]=
(1/2\hbar )\int_{t_i}^{t_f}\int_{t_i}^{t_f}dtdsN(t-s)y(t)y(s). \eqno(30c)
$$
The central Eqs.(30) are {\it rigorously exact} for linear passive damping
due to the bath when the path integral Eq.(15) is employed for the
time development of the density matrix.

The lagrangian Eq.(2) can now be viewed as the approximation
to Eq.(30b) with $F_y^{ret}=R\dot{y}$ and $F_x^{adv}=-R\dot{x}$.
The classical constraint $y=0$ occurs because nonzero $y$ yields an
``unlikely process'' in view of the large imaginary part of the action
(in the classical ``$\hbar \rightarrow 0$''limit) implicit in Eq.(30c).
On the contrary, at quantum level nonzero $y$ may allow quantum
noise effects arising from the imaginary part of the action.

We are glad to aknowledge useful discussions with E. Celeghini and M.Rasetti.

\vfill
\eject

\bigskip
\centerline{\bf References}
\bigskip

\item{1.} H. Bateman, {\it Phys. Rev.} {\bf 38} (1931), 815

\item{2.} H. Feshbach and Y. Tikochinsky, {\sl Transact. N.Y.
Acad. Sci.} {\bf 38} (Ser. II) (1977) 44

\item{3.} E. Celeghini, M. Rasetti and
G. Vitiello, {\it Annals of Phys. (N.Y.)} {\bf 215} (1992), 156

\item{4.} Y. Takahashi and H. Umezawa, {\it Collective Phenomena}
{\bf 2} (1975), 55;

\item{} H.Umezawa, {\it Advanced field theory: micro,
macro and thermal concepts} (American Institute of Physics, N.Y. 1993)

\item{5.} J. Schwinger, {\it J. Math. Phys.} {\bf 2} (1961), 407

\item{6.} R.P. Feynman and F.L. Vernon, {\it Annals of Phys. (N.Y.)}
{\bf 24} (1963) 118

\vfill
\eject
\bye